\documentclass[preprint2]{aastex7}
\usepackage{caption,subcaption}
\usepackage[english]{babel}
\usepackage{graphicx}
\usepackage{amsmath}
\usepackage{natbib}
\usepackage{gensymb}
\usepackage{multirow}
\usepackage{xcolor}
\usepackage{soul}
\usepackage{cancel}

\begin{document}
\title{Spectral Analysis of the 2019 and 2022 Outbursts of SAX J1808.4-3658}

\author[0000-0001-7097-3615]{Katherine Bruce}
\affiliation{Department of Physics, Montana State University, 
Bozeman, MT 59717, USA}
\email{katherinebruce@montana.edu}

\author{Sachiko Tsuruta}
\affiliation{Department of Physics, Montana State University, 
Bozeman, MT 59717, USA}
\affiliation{Kavli Institute for the Physics and Mathematics 
of the Universe (WPI), The University of Tokyo, Kashiwa, Chiba 
277-8583, Japan}
\email{tsuruta@montana.edu}

\author{Andrew C. Liebmann}
\affiliation{Department of Physics, Montana State University, 
Bozeman, MT 59717, USA}
\email{bloodcorba@earthlink.net}

\author{Marcus Teter}
\affiliation{RTX, 16510 E. Hughes Dr., 
Aurora, CO, 80011, USA}
\affiliation{Department of Physics, Montana State University, 
Bozeman, MT 59717, USA}
\email{marcus.a.teter@gmail.com}

\accepted{26 Nov 2025}

\begin{abstract}

The accreting millisecond pulsar SAX J1808.4-3658 went into outburst from July to November in 2019 and August to October in 2022, which were observed by \textit{NICER} and \textit{NuSTAR}. In this paper, we first present the light curve for both outbursts using \textit{NICER} data. Several thermonuclear bursts occurred during these outbursts. We analyze the evolution of the spectra of two thermonuclear bursts that took place during the 2019 \textit{NuSTAR} observation. We proceed by analyzing the combined broad-band spectrum using \textit{NICER} and \textit{NuSTAR} for the first time for this source. We jointly modeled the combined quiescent spectra of both outbursts with a self-consistent reflection component. In our best-fit model, we find evidence of reflection, consistently constrain the inclination to 72\degree$^{+1\degree}_{-4\degree}$\, considering this reflection, and identify a 1 keV feature during persistent emission.

\end{abstract}

\section{Introduction} \label{sec:intro}

The accreting millisecond pulsar SAX J1808.4-3658 (hereafter SAX J1808) was the first of its kind to be detected \citep{wijnands}. This neutron star exists alongside a brown dwarf companion (M$_{\text{c}}\approx 0.05 M_{\odot}$) in a low-mass X-ray binary (LMXB) and engages in outbursts frequently, every 2-4 years \citep{bildsten, bult}. Outbursts have been detected by the \textit{Rossi X-Ray Timing  Explorer (RXTE)}, the \textit{Chandra X-Ray Observatory}, \textit{Suzaku}, \textit{XMM-Newton}, the \textit{Nuclear Spectroscopic Telescope Array (NuSTAR)}, and the \textit{Neutron Star Interior Composition Explorer (NICER)} and occurred in 1998, 2002, 2005, 2008, 2011, 2015, 2019, and 2022 \citep{zand,markwardt02,rupen,markwardt08,altamirano, sanna}. This source is well-studied and is known to have a spin frequency of 401 Hz and an orbital period of 2.1 hours \citep{wijnands,chakrabarty}. This source has been classified as an atoll source with a distance of $\sim$3.5 kpc \citep{straaten,galloway08}. This source does not behave like a usual atoll source, primarily inhabiting the island (hard) state, rarely shifting into the extreme island (hardest) and lower left banana (soft) states \citep{bult2015}. \cite{goodwin2019} used a new burst mechanism model alongside Multi-INstrument Burst ARchive (MINBAR) data to infer a neutron star mass of 0.9-1.9 M$_{\odot}$ and a 10.0-13.1 km radius.

The increased luminosity from X-ray outbursts allows us to observe and study SAX J1808 in more detail. During outbursts, the hot corona emits X-rays that illuminate the accretion disk below \citep{fabian2000}. For pulsars, this accretion disk is most likely truncated by a few gravitational radii from the stellar surface due to the magnetosphere \citep{romanova}. Much of the power-law component escapes directly from the corona, while some of its photons illuminate the disk near the inner boundary. This results in reflection features in the spectra, such as the Fe K$\alpha$ emission line and a hard X-ray hump near 20-40 keV \citep{fabian2000}. The hard hump is caused by both photoelectric absorption of low-energy photons and Compton down-scattering of high-energy photons. The Fe K$\alpha$ emission line is an intrinsically narrow line, with two components at 6.404 keV and 6.391 keV for neutral iron. The ionization state of iron can shift this emission line as high as 6.97 keV \citep{fabian2000, osterbrock}. Due to gravitational effects from the proximity to the compact object, these emission lines are usually broadened and skewed by both the Doppler effect and gravitational redshift \citep{fabian2000}. 

While the reflection phenomenon was originally identified in active galactic nuclei, the Fe K$\alpha$ emission line has been identified in neutron stars in LMXBs \citep{white}. For SAX J1808, this feature was first detected in the 2008 outburst \citep{papitto09}. \cite{disalvo} analyzed the \textit{NuSTAR} and \textit{XMM-Newton} observation of this source's 2015 outburst and identified both the Fe K$\alpha$ emission line and the Compton hump, along with the presence of three additional emission lines in the 2-5 keV range of the \textit{XMM-Newton} data. \cite{bult} analyzed one \textit{NICER} observation from the 2019 outburst, which captured a particularly bright thermonuclear burst. Their analysis identified the 6.7 keV Fe K$\alpha$ emission line and a 1.0 keV emission feature. This 1 keV emission feature, which is well known to exist in X-ray binaries, could be a blend of emission lines, some arising from the reflection Fe L complex \citep{chakraborty,ludlam2018}. 

In addition to the outbursts, SAX J1808 also frequently engages in thermonuclear (type-I) X-ray bursts \citep{galloway, bult}. Accreted mass builds up on the surface, raising pressure and temperature. This causes unstable nuclear burning of pure helium, eventually igniting a thermonuclear eruption lasting until the accreted material is burned \citep{galloway08}. The characteristic cycle for these bursts is a fast rise and exponential decay \citep{chen}. These bursts last on the order of seconds, and recurrence depends on mass accretion rate, usually on the order of hours. In some cases, enough energy is released to reach and exceed the Eddington limit, causing the atmosphere to expand due to radiation pressure \citep{kuul}. SAX J1808 is known to exhibit this so-called photospheric radius expansion \citep{galloway08}. \cite{speicher2022} recently concluded that type-I bursts and changes in disk geometry can affect the strength of the emission lines that make up the soft excess, which may explain the variation in strength and centroid of the 1 keV feature.  

With the impact of the thermonuclear bursts on the soft excess and observed 1 keV line features uncertain, we study the recent quasi-simultaneous \textit{NICER} and \textit{NuSTAR} observations with the hope of understanding spectral decomposition better. In particular, we aim to investigate potential thermonuclear bursts in these recent observations and confirm that the 1 keV emission feature can be explained in part by reflection from the disk during non-burst emission.

Our paper is structured as follows: In Section \ref{sec:outburst}, we describe the creation of the \textit{NICER} and \textit{NuSTAR} light curves and spectra. Section~\ref{sec:lc} places the observations in the context of the long-term light curves and details the identification of two thermonuclear bursts in the 2019 NuSTAR observation. The modeling of their time-resolved burst spectra is described in Section \ref{burst}. Finally, in Section \ref{sec:analysis}, we present our modeling and analysis of the broad-band spectra using combined \textit{NICER} and \textit{NuSTAR} spectral data. The findings are then discussed in Section \ref{sec:discussion}. 

\section{Observations and Data Reduction} \label{sec:outburst}

Our focus in this paper are the 2019 and 2022 outbursts. During these outbursts, SAX J1808 was observed with \textit{NuSTAR} on 2019 August 10 for 41 ks (ObsID: 90501335002) and on 2022 August 22 for a length of 107 ks (ObsID: 80701312002). The data were processed using the standard pipeline of \texttt{NuSTARDAS} (NuSTAR data analysis software) v 2.1.4, specifically using the tasks \texttt{nupipeline} and \texttt{nuproducts} \citep{harrison}. After initially cleaning the event file, we created the source and background regions and extracted the spectra and light curves. For both observations, the source region was selected to be a circular region with 120 arcmin radius, and the background region was selected to be a circular region with 170 arcmin radius, far from the source region. The resulting spectra of both focal plane modules A and B were modeled simultaneously alongside the \textit{NICER} spectra (see next section) in the energy range 3.0-60.0 keV, where the upper limit is selected to be where the source spectrum crossed the background. 

\textit{NICER} captured 63 usable exposure observations between 2019 July 30 and 2019 November 8 and 53 observations between 2022 August 19 and 2022 October 31. Each of these observations was processed using the HeaSOFT v6.34 \textit{NICER} Data Analysis System (\texttt{NICERDAS}) v13 software, using the standard processing pipeline \texttt{nicerl3-lc} to extract light curves with a bin size of 1.0 seconds \citep{blackburn}. For the spectral analysis, the \textit{NICER} observations taken closest to \textit{NuSTAR} were captured on 2019 August 10, lasting for 5.4 ks (ObsId:2584010103), and on 2022 August 23, for 10.6 ks (ObsId:5050260105). These data were processed using  \texttt{nicerl3-spect} to extract the relevant spectral data and account for the systematic error of 1.5\%. The background for \textit{NICER} has been estimated using the SCORPEON v23 \footnote{\url{https://heasarc.gsfc.nasa.gov/docs/nicer/analysis_threads/scorpeon-xspec/}}background model. The spectral energy range used for \textit{NICER} was 0.7-10.0 keV, with the lower limit chosen to eliminate the low energy noise occurring in data taken during orbit day. All spectra were binned to a minimum of 20 counts per bin for modeling. 

\section{Analysis and Results}

\subsection{Light Curve}\label{sec:lc}

\begin{figure*}
  \centering
  \begin{subfigure}{0.45\textwidth}
    \centering
    \includegraphics[width=\linewidth]{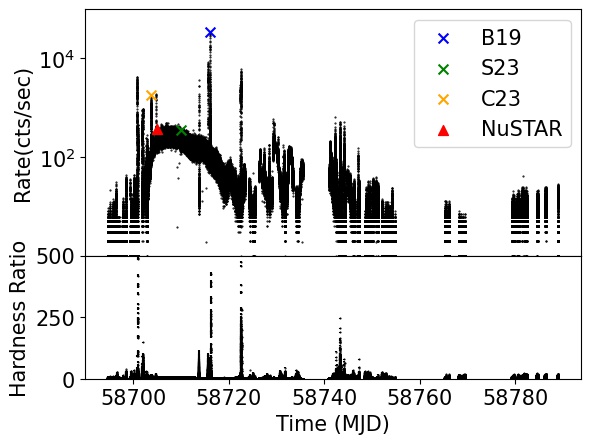}
    \caption{2019 July 30 to 2019 November 8}
    \label{fig:19LC}
  \end{subfigure}%
  \quad
  \begin{subfigure}{0.45\textwidth}
    \centering
    \includegraphics[width=\linewidth]{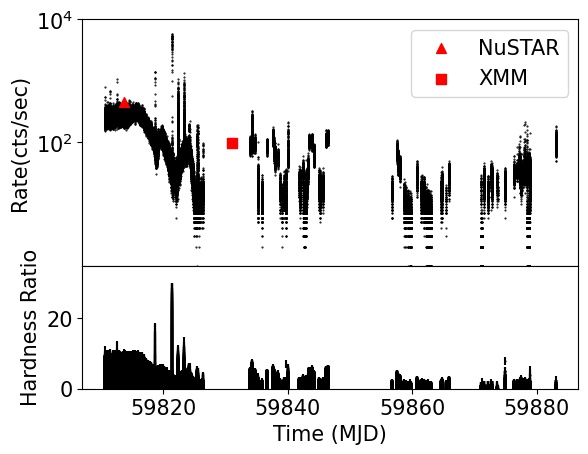}
    \caption{2022 August 19 to 2022 October 31}
    \label{fig:22LC}
  \end{subfigure}
  \caption{Top Panel: Light curves created using \textit{NICER} data. Binned by second to illustrate any short bursts occurring. Bottom Panel: Hardness ratio of (6-12 keV)/(0.3-6 keV). B19: \cite{bult}. S23: \cite{sharma23}. C23: \cite{casten23}. Note: \textit{NuSTAR} and \textit{XMM-Newton} rates converted to \textit{NICER} rates using WebPIMMS.}
  \label{fig:LC}
\end{figure*}

Figure \ref{fig:LC} shows the \textit{NICER} one second light curves of SAX J1808 during its 2019 (left panel) and 2022 (right panel) outbursts. The hardness ratio shown in the bottom panel is chosen to be (6-12 keV)/(0.3-6 keV). From the 2019 light curves, we learn that the system reached a maximum observed count rate of 33661 ct s$^{-1}$ on 2019 August 21 (ObsID: 2584010501). This observation, which coincides with an X-ray burst, was analyzed in \cite{bult}. This observation and other studies of this outburst by \cite{sharma23} and \cite{casten23} have also been marked in Figure \ref{fig:19LC}. At least 11 other bursts have been identified during this outburst. The \textit{NuSTAR} observation averaged a rate of 15 ct s$^{-1}$ during its observation, and, using the HEASARC WebPIMMS tool \footnote{\url{https://heasarc.gsfc.nasa.gov/cgi-bin/Tools/w3pimms/w3pimms_pro.pl}}, the \textit{NICER} count rate for this observation is approximately 366 ct s$^{-1}$. Shown in Figure \ref{fig:22LC}, the 73-day light curve of the 2022 outburst reaches a maximum observed rate of 5775 ct s$^{-1}$ on 2022 August 30 (ObsID:5574010108). About 8 bursts occurred during this outburst. The \textit{NuSTAR} observation had an average rate of 17 ct s$^{-1}$, and the WebPIMMS tool estimated the corresponding \textit{NICER} count rate to be approximately 456 ct s$^{-1}$. For the \textit{XMM-Newton} observation, the EPIC/pn detector measured an average rate of 66 ct s$^{-1}$, and the webPIMMS tool estimated a \textit{NICER} rate of 98 ct s$^{-1}$. Both \textit{NuSTAR} observations were taken right before the peak of the outburst. While the 2019 outburst has the brighter thermonuclear burst, the \textit{NuSTAR} observation and its corresponding \textit{NICER} observation are brighter in 2022 than in 2019. 

Analysis of the 2019 \textit{NuSTAR} observation reveals that two thermonuclear bursts occurred. The first started 32280 seconds into the observation, and the second began 72010 seconds into the observation, occurring approximately 11 hours apart. 

\subsection{Thermonuclear Bursts}\label{burst}

The light curves of the two thermonuclear bursts showed the typical fast-rise exponential-decay burst behavior. To further explore the nature of these bursts, they were split into 4 10-second intervals to best bin the data with a signal-to-noise ratio of 3 and to study the evolution of the spectra throughout the burst. The four intervals represent one rise, one peak, and two decline sections. The energy range used for the \textit{NuSTAR} spectra of each interval began at the soft cutoff at 3.0 keV and ended when the signal crossed the background, ranging between 8.0- 15.0 keV. During a thermonuclear burst, the X-ray spectrum has been observed to be dominated by a blackbody component due to the eruption occurring on the star's surface. Each spectrum was therefore modeled in \texttt{XSPEC} v12.14.1 with absorption and a blackbody component, with the hydrogen absorption column, interstellar oxygen abundance, and interstellar iron abundance frozen to the best-fit values of $0.23\times 10^{22}$ cm$^{-2}$, 1.6 $_{-0.38}^{+0.63}$ solar units, and 1.4 $_{-0.98}^{+0.86}$ solar units from the analysis of the quiescent spectrum, respectively (see Section \ref{sec:analysis}). Such a model describes the spectrum well, and we found no evidence of other components in any time interval. Each model resulted in a reduced $\chi^{2}$ of 1.00-1.36. Residuals from each burst and
spectra are shown in the Appendix. No other features or systematic errors were detected. The errors were determined at 90\% from a Markov Chain Monte Carlo (MCMC) with a chain length of $1\times10^{3}$, 10 walkers, and a burn length of 50.  

\begin{figure*}
  \centering
  \begin{subfigure}{0.45\textwidth}
    \centering
    \includegraphics[width=\linewidth]{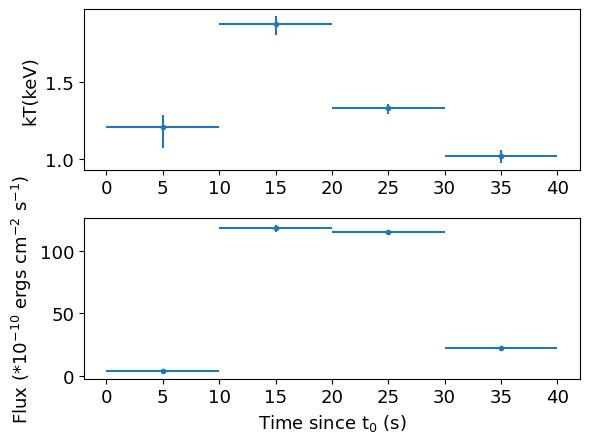}
    \caption{Burst 1, starting at t$_{0}$=32280 s}
    \label{fig:burst1}
  \end{subfigure}
  \quad
  \begin{subfigure}{0.45\textwidth}
    \centering
    \includegraphics[width=\linewidth]{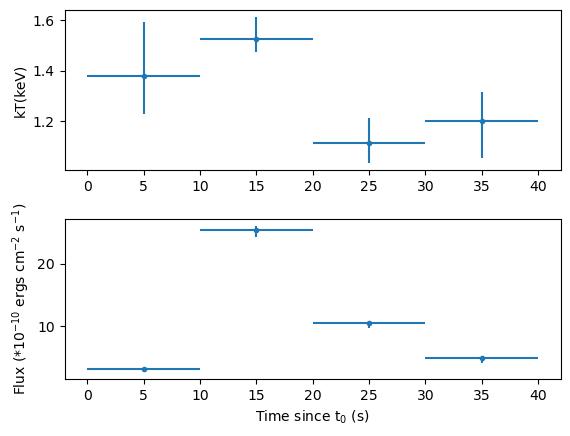}
    \caption{Burst 2, starting at t$_{0}$=72010 s}
    \label{fig:burst2}
  \end{subfigure}
  \caption{Temperature and flux throughout the thermonuclear bursts occurring in the 2019 \textit{NuSTAR} observation. Data modeled with photoelectric absorption (including interstellar oxygen and iron abundance) and blackbody components. Bursts display the commonly found fast-rise exponential-decay burst profile.}
  \label{fig:burst}
\end{figure*}

\begin{table*}
    \captionsetup{justification=centering}
    \caption{Best-fit parameters for the 2 thermonuclear bursts occurring in the 2019 \textit{NuSTAR} observation. The blackbody radius was calculated using a distance of ~3.5 kpc \citep{galloway08}. Top: First burst, beginning at t$_{0}$= 32280s. Bottom: Second burst, beginning at t$_{0}$ = 72010 s}
     \label{tab:burstparams}
    \begin{tabular}{lcccc}
        \hline
        Section  & Blackbody Temp  & Blackbody Radius & Flux& $\chi^{2}$/dof  \\
         & (keV) & (km) & (*$10^{-10}$ ergs cm$^{-2}$ s$^{-1}$)  & \\
        \hline
	Rise &  1.2 $_{-0.14} ^{+0.07} $ & 1.8 $_{-0.3} ^{+0.6} $& 4.0 $_{-0.6} ^{+0.2} $& 23/18 \\
	Peak & 1.9  $_{-0.07} ^{+0.05} $&  3.5$_{-0.2} ^{+0.3} $&118 $_{-3} ^{+2} $& 265/227 \\
	Decline 1 & 1.3 $_{-0.03} ^{+0.03} $ & 7.4 $_{-0.4} ^{+0.5} $& 115 $_{-2} ^{+1} $ &  230/231 \\
	Decline 2 & 1.0 $_{-0.04} ^{+0.04} $ & 6.2$_{-0.7} ^{+0.6} $& 22.2 $_{-1.5} ^{+0.6} $& 91/79 \\
	\hline
	Rise &  1.51 $_{-0.24} ^{+0.16} $ &  1.1 $_{-0.4} ^{+0.2} $&  3.2 $_{-0.5} ^{+0.1} $ &  29/17 \\
	Peak & 1.52  $_{-0.05} ^{+0.09} $ & 2.7$_{-0.3} ^{+0.2} $&25.4 $_{-1.1} ^{+0.7} $& 111/104 \\
	Decline 1 & 1.11 $_{-0.08} ^{+0.10} $ & 3.7$_{-0.7} ^{+0.8} $ &10.5 $_{-0.8} ^{+0.2} $&  53/45 \\
	Decline 2 & 1.2  $_{-0.15} ^{+0.12} $& 2.2$_{-0.5} ^{+0.7} $& 4.9 $_{-0.8} ^{+0.1} $& 24/22  \\
         \hline
    \end{tabular}
\end{table*}
The temporal evolution of the unabsorbed flux and blackbody temperature is visualized in Figure \ref{fig:burst}. The best-fit parameters are shown in Table \ref{tab:burstparams}. We observe changes in both blackbody temperature, blackbody radius, and unabsorbed flux through the evolution of the burst. The evolution of the temperature and flux behaves as expected, rising quickly, peaking, and declining exponentially. This trend is much clearer in the first burst, being the brighter of the two. The measured blackbody radius is largest around the time of the peak flux or shortly thereafter in both bursts, then it shrinks again as the flux declines.

\subsection{Broad-band Spectral Analysis} \label{sec:analysis}

\begin{figure*}
  \centering
  \begin{subfigure}{0.48\textwidth}
    \centering
    \includegraphics[width=\linewidth]{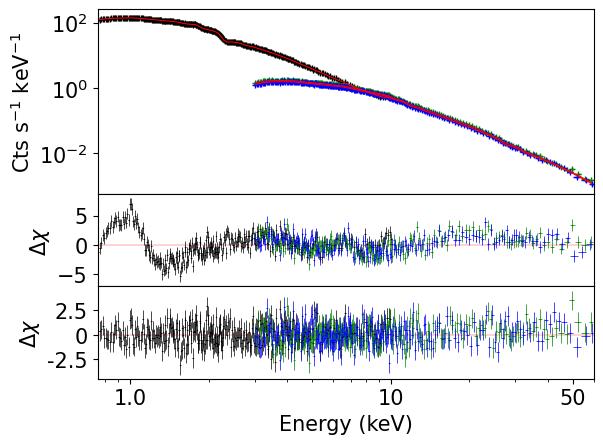}
    \caption{2019 Spectra}
    \label{fig:19spect}
  \end{subfigure}%
  \quad
  \begin{subfigure}{0.48\textwidth}
    \centering
    \includegraphics[width=\linewidth]{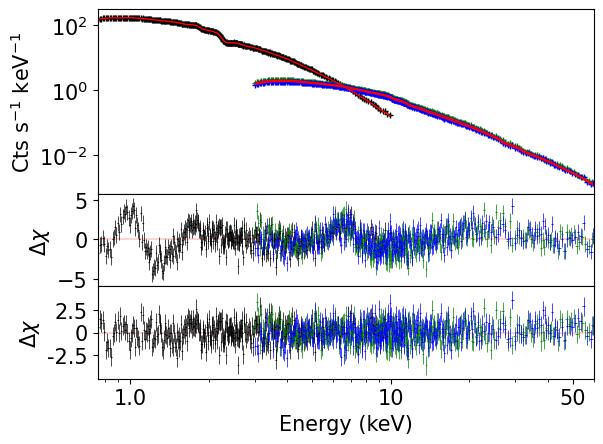}
    \caption{2022 Spectra}
    \label{fig:22spect}
  \end{subfigure}
  \caption{Top: Broadband \textit{NICER} (black) and \textit{NuSTAR} (blue) spectrum and best-fit model (red) of SAX J1808. (a) depicts the 2019 outburst and (b) the 2022 outburst. Model includes photoelectric absorption (including interstellar iron and oxygen abundance), a thermally Comptonized primary continuum, a blackbody, a Gaussian line, and a reflection component. Middle: Residuals in units of sigma of the data to the model without the Gaussian line at 1 keV and the reflection component. Here, the 1 keV line, the iron line, and the hard hump appear. Bottom: Residuals in units of sigma of the data to the best-fit model, including the Gaussian line and reflection component.}
  \label{fig:spect}  
\end{figure*}

\begin{table*}
    \captionsetup{justification=centering}
    \caption{Best-fit parameters for the combined broad-band spectral analysis of the 2019 and 2022 outbursts of SAX J1808 using \textit{NICER} and \textit{NuSTAR}. The model consists of photoelectric absorption  (including interstellar iron and oxygen abundance), a blackbody, a Gaussian emission line, a thermally Comptonized primary continuum, and the reflection component. The blackbody radius was calculated using a distance of ~3.5 kpc \citep{galloway08}. The iron abundance is in units of solar iron abundance.}
     \label{tab:params}
    \setlength{\tabcolsep}{10pt}
    \renewcommand{\arraystretch}{1.55}
    \centering
    \begin{tabular}{llcc}
        \hline
        Component & Parameter & 2019 & 2022  \\
        \hline
         tbfeo & N$_{\text{H}}$ (*10$^{22}$ cm$^{-2}$) & 0.230 $_{-0.0322}^{+0.039}$ & 0.173 $_{-0.0208}^{+0.0224}$\\
         tbfeo & O & 1.57 $_{-0.377}^{+0.629}$ & 1.34 $_{-0.356}^{+0.372}$  \\
         tbfeo & Fe & 1.36 $_{-0.980}^{+0.857}$ & 3.03 $_{-0.567}^{+0.800}$  \\
         bbodyrad & kT$_{\text{bb}}$ (keV) & 0.558 $_{-0.0126}^{+0.009}$& 0.630 $_{-0.0135}^{+0.011}$\\
         bbodyrad & R$_\text{bb}$ (km) & 3.1 $_{-0.15}^{+0.26}$ & 2.4 $_{-0.11}^{+0.17}$ \\
         gaussian & LineE (keV) & 0.884  $_{-0.0158}^{+0.0215}$ & 0.988  $_{-0.041}^{+0.017}$ \\
         gaussian & Intensity ($\times 10^{-2}$ ph cm$^{-2}$ s$^{-1}$)& 1.23  $_{-0.311}^{+0.760}$ & 0.41 $_{-0.099}^{+0.243}$\\
         gaussian & Width (keV) & 0.139  $_{-0.0160}^{+0.0311}$ & 0.107 $_{-0.0158}^{+0.0349}$ \\
         relxillCp & $\Gamma$ & 1.96 $_{-0.006}^{+0.023}$ & 1.99 $_{-0.078}^{+0.010}$\\
         relxillCp & kT$_{\text{e}}$ (keV) & \multicolumn{2}{c}{74.2 $_{-11.32}^{+21.10}$}\\
         relxillCp & kT$_{\text{seed}}$ (keV)  & \multicolumn{2}{c}{0.01 (fixed)} \\
         relxillCp & Index & 2.05  $_{-0.247}^{+0.197}$ & 2.21 $_{-0.111}^{+0.228}$\\
         relxillCp & R$_{\text{in}} (GM/c^2) $ & 16.1 $_{-5.20}^{+3.92}$ & 12.3  $_{-3.42}^{+7.22}$\\
         relxillCp & Incl (deg)  &\multicolumn{2}{c}{72.3 $_{-3.61}^{+0.56}$}\\
         relxillCp & LogN (cm$^{-3}$) & 19.1 $_{0.17}^{+0.34}$& 17.3  $_{-0.64}^{+0.32}$\\
         relxillCp & R$_{f}$ & 0.329  $_{0.070}^{+0.079}$ & 0.527 $_{-0.1066}^{+0.0491}$ \\
         relxillCp & Fe abund & \multicolumn{2}{c}{1.0 (frozen)}  \\
         relxillCp & log$\xi$ & 1.65 $_{-0.399}^{+0.227}$& 1.82 $_{-0.199}^{+0.234}$ \\      
         relxillCp & Norm (*10$^{-3}$)& 1.9 $_{-0.03}^{+0.07}$ & 2.2 $_{-0.03}^{+0.06}$ \\
         \hline
         Total & $\chi^{2}$/DOF & 2569 / 2785 & 3039 / 3291 \\
         Total & $\chi^{2}$/DOF & \multicolumn{2}{c}{ 5608 / 6076 } \\
         \hline
    \end{tabular}
\end{table*}

\begin{figure*}
  \centering
  \begin{subfigure}{0.45\textwidth}
    \centering
    \includegraphics[width=\linewidth]{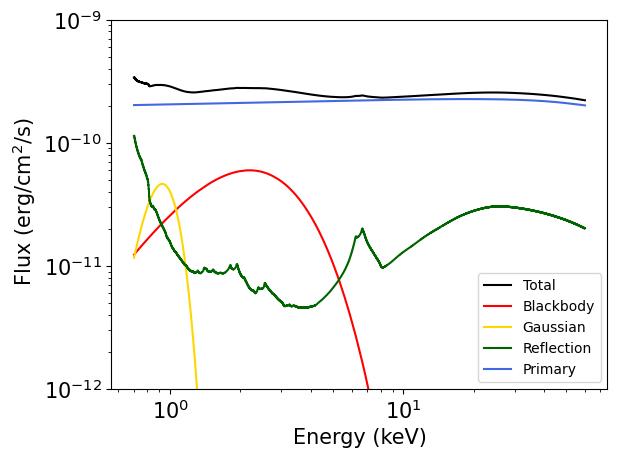}
    \caption{2019}
    \label{fig:2019comp}
  \end{subfigure}
  \begin{subfigure}{0.45\textwidth}
    \centering
    \includegraphics[width=\linewidth]{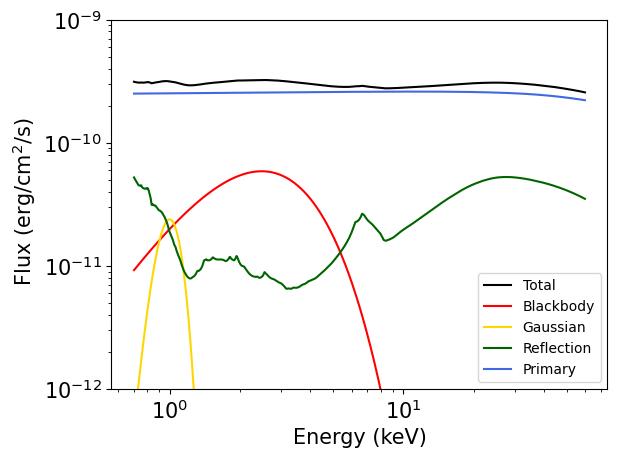}
    \caption{2022}
    \label{fig:2022comp}
  \end{subfigure}
  \caption{Relative strength of the components from the best-fit model using \textit{NICER} and \textit{NuSTAR} for 
  the 2019 and 2022 outbursts of SAX J1808 from Section \ref{sec:analysis}.}
  \label{fig:continuum}
\end{figure*}

After modeling the spectral evolution throughout the burst in the last section, this section focuses on the modeling of the broad-band spectra outside of the bursts. Our modeling was done in \texttt{XSPEC} v12.14.1 and \texttt{PyXspec} v2.1.4 \citep{arnaud}. Both \textit{NuSTAR}'s FPMA and FPMB were modeled alongside the \textit{NICER} data. Each outburst was initially modeled with photoelectric absorption (\texttt{TBFEO}), a blackbody (\texttt{BBODYRAD}), and a thermally Comptonized continuum (\texttt{NTHCOMP}) \citep{wilms,zdziarski,zycki}. \cite{goodwin2020b} reports a disk temperature of 1-2 eV at the onset of X-ray detection on August 6th in 2019. Considering the 2022 outburst was taken at a similar point during the outburst, it is reasonable to assume the disk is also too soft in 2022 to be properly modeled. In each model, the photoelectric absorption and the interstellar oxygen and iron abundances were free to vary, rather than being frozen to a previously reported value. Both displayed a significantly improved result with the inclusion of a blackbody component. Emission line residuals still appeared near 1 and 6.5 keV, and a broad hump around 20-40 keV. 
The result further improved with the inclusion of a Gaussian line and the self-consistent reflection model \texttt{relxillCp} \citep{garcia}. The Cp flavor of the \texttt{relxill} family uses a thermally Comptonized primary continuum and a broken power law for the disk emissivity and fixes the seed photon temperature to 0.01 keV. For these models, the disk emissivity was treated as a power law function of the radius with the index left free to vary. The outer radius of the disk was set to 1000 gravitational radii and frozen. The spin was estimated to $\approx$ 0.3 using the spin frequency of 401 Hz \citep{galloway08}. 

For the 2019 spectrum, the model results in an absorption of $\approx0.24^{+0.05}_{-0.06}\times 10^{22}$ cm$^{-2}$, while it gives 0.17$_{-0.04}^{+0.03}\times 10^{22}$ cm$^{-2}$ for the 2022 spectrum. The Gaussian emission line was found at 0.88 $_{-0.1}^{+0.02}$ keV in 2019 and at 0.98 $_{-0.07}^{+0.02}$ keV in 2022. In the best-fit for the 2022 model, we obtain an inclination of 77\degree$_{-5\degree}^{+3\degree}$\, and an inner disk radius of 14 $_{-4}^{+12}GM/c^2$. The 2019 model settled on a lower inclination and higher inner disk radius, 72\degree$_{-12\degree}^{+1\degree}$\, and 16 $_{-6}^{+18}GM/c^2$, respectively. 

The large difference in the inclination between years led us to believe one or both of these results are incorrect and motivated us to model the spectra from both outbursts together. Our combined analysis of the 2019 and 2022 outbursts allows us to force self-consistency on static parameters, specifically inclination. All other parameters may differ between years and were left free to vary. This was accomplished using the Tempest High Performance Computing System at Montana State University. These best-fit parameters are shown in Table \ref{tab:params}. The spectra for these best-fit parameters are shown in Figure \ref{fig:spect}. The middle panel excludes the Gaussian line and reflection components. The relative strengths of each component in the best-fit model are shown in Figure \ref{fig:continuum}. All spectra were modeled and errors were found using an MCMC with 400 walkers, a 3$\times 10^{6}$ chain length, and a 4$\times 10^{4}$ burn length. Errors are reported at a 90\% confidence interval. For the 2019 model, the $\chi^{2}$ decreased by ~600 with -9 degrees of freedom with the inclusion of the reflection and Gaussian components. The $\chi^{2}$ in 2022 decreased by ~570 with -9 degrees of freedom with the inclusion of the reflection and Gaussian components.    

The Gaussian emission line near 1 keV is detected in both outbursts, but is found at 0.90 $_{-0.02}^{+0.02}$ keV in 2019 and at 0.99 $_{-0.02}^{+0.04}$ keV in 2022. Many parameters, such as emissivity index, $\Gamma$, and blackbody temperature, agree fairly well between outbursts and fall into accepted ranges. For both years, the inner radii settled to values on the larger side of what is considered acceptable, but were not well constrained by the models. The inclination between 68-73\degree\, agrees with previous studies, which identify this system as having a high inclination (see \cite{disalvo} for further discussions). 

\section{Discussion} \label{sec:discussion}
In this paper, we created light curves using \textit{NICER} observations of SAX J1808 in 2019 and 2022 and found several thermonuclear bursts in these observations. We also identified two bursts in the 2019 \textit{NuSTAR} observation, and performed time-resolved spectral analysis to study the evolution. Finally, we modeled the broad-band spectrum for the first time using \textit{NICER} and \textit{NuSTAR} data. In this model, we found evidence of reflection during both outbursts, constrained the inclination to 72 $_{-3.6}^{+0.6}$\degree\, with the reflection component, and identified the 1 keV line during non-burst emission.

The 1 keV line feature is well known to exist in X-ray binaries and has previously been identified in sources like Serpens X-1, Cygnus X-2, Hercules X-1, and HETE J1900.1-2455 \citep{chakraborty,pappitto2013}. Because this feature is believed to be a blend of emission lines, including the Fe L complex from reflection, the relative strengths of the lines varying is likely causing the shift in centroid \citep{chakraborty, ludlam2018}. Simulations of a Type-I thermonuclear burst in an LMXB by \cite{speicher2022} suggest that these bursts are likely to ionize and heat the disk, altering the strength of the soft excess as a result.

We identified an emission line at 0.90 $_{-0.02}^{+0.02}$ keV with an intensity of 0.012 $_{-0.003}^{+0.008}$  ph cm$^{-2}$ s$^{-1}$, approximately 20 hours before the bursts in the 2019 \textit{NuSTAR} observation. In 2022, we find this line at 0.99 $_{-0.02}^{+0.04}$ keV, with a much smaller intensity of 0.004 $_{-0.001}^{+0.002}$ ph cm$^{-2}$ s$^{-1}$. Between these outbursts, the reflection fraction increases from 0.33  $_{0.07}^{+0.08}$ to 0.53 $_{-0.10}^{+0.05}$. This variance in reflection fraction may contribute to the shift in central energy, but this is likely not the case. We expect that an increase in the reflection fraction would correspond to a brighter feature. This does not exclude the possibility of reflection affecting this feature, but we do not find conclusive evidence here. Similarly, the ionization parameter log $\xi$, which varies from 0.0 to 4.7, increases from 1.7 $_{-0.4}^{+0.2}$ to 1.8 $_{-0.2}^{+0.2}$ between the two epochs. This change in ionization is likely correlated with the shift in central energy of this feature. The theory that this feature is a blend of soft emission lines aligns with this explanation.

In fact, \cite{bult} analyzed the brightest thermonuclear burst of SAX J1808 in 2019 and found an emission line at 1.05 keV, with an intensity of 0.27 ph cm$^{-2}$ s$^{-1}$. This observation took place 12 days after the \textit{NICER} observation analyzed in this paper. The much higher intensity makes sense considering the brightness of that thermonuclear burst. \cite{sharma23} identified this feature during persistent emission at 0.86 keV, with an intensity of 0.002 ph cm$^{-2}$ s$^{-1}$. This observation used \textit{AstroSAT} and occurred 4 days after the \textit{NuSTAR} observation in 2019. Finally, \cite{casten23} analyzed a \textit{NICER} observation occurring one day before the observation analyzed in this paper. They found this feature at 0.99 keV before the thermonuclear burst and at  0.96 keV during the burst. They did not find this feature in the post-burst spectrum. Between these analyses, the central energy shifts by 0.2 keV. Such a vast variation in the central energy of this feature is not unusual. \cite{ludlam2018} finds this feature in Serpens X-1 at 1.22 keV and later \cite{hall25} identifies this feature at 0.87 keV.

There are other interesting emission lines to be analyzed in SAX J1808. \cite{disalvo} reported three soft emission lines in a 2015 \textit{XMM-Newton} observation at 2.7, 3.3, and 4.0 keV, thought to be Sulfur XVI, Argon XVIII, and Calcium XIX-XX, respectively. In our analysis of the 2019 and 2022 \textit{NICER} and \textit{NuSTAR} observations, these lines were not identified. While the nature of these lines is still unknown, their presence may be explained by the unusually soft spectral state of this source in the 2015 \textit{XMM-Newton} observation. We emphasize that this source should be observed with instruments with higher spectral resolution, e.g., with \textit{XRISM}, to further investigate the origin of these lines. 

Modeling the combined broadband spectrum with the blackbody and reflection components increased the quality of our results significantly for the 2022 spectra, giving good evidence that these components are occurring in the system. Forcing certain parameters to be self-consistent between outbursts, such as inclination, also resulted in an increased quality of our results. This included a reasonable inclination of 69-73$^{\circ}$, which falls within the recently modeled range of 67-73$^{\circ}$ found by \cite{goodwin2019}. Considering that \cite{bult} identified an accretion disk in their analysis, we considered including the accretion disk in our model. However, \cite{goodwin2020b} projects the disk temperature to be on the order of a few eV about two weeks before the \textit{NuSTAR} observation. Considering their projected viscous time-scale of 4-12 days, we do not expect the accretion disk to appear in X-ray. The 2022 observation was taken at a similar point during the outburst and is likely also emitting below X-ray in this observation. 

Studies of previous outbursts have found upper limits for the inner disk radii of around 12 R$_{g}$ during the 2008 and 2015 outbursts, and this is consistent with our findings for the 2019 and 2022 outbursts \citep{papitto09,disalvo}.

Assuming a 1.4 M$_{\odot}$ mass star, the inferred radius ranged from roughly 20 to 40 km for both outbursts. We expect the accretion disk to lie within the co-rotation radius a few days before the outburst peak when the \textit{NuSTAR} observations were taken. Considering the co-rotation radius of 31 km estimated by \cite{patruno}, these radii are at the larger end of what would be expected for an accreting neutron star. The large error range does not allow us to conclude with certainty whether the inner disk radius lies within or outside of the co-rotation radius.
SAX J1808 is well known to exhibit reflaring behavior, where the source erratically re-brightens after the outburst peak \citep{wijnands2001,patruno2009,baglio2020}. During this period, the accretion disk is theorized to retreat and outflow rather than accrete \citep{patruno}. \cite{baglio2020} found the source to be in an island/hard state for the 2019 outburst and that the reflaring regime begins approximately 13 days after the X-ray peak (17 days before the \textit{NuSTAR} observation). \citep{asquini} finds further evidence of this retreat, reporting the inner disk radius at $\approx$5100 km (over 10$^{3}$ R$_{g}$) 3 weeks after the 2019 outburst. Explanations suggested for this changing radius include changes in accretion geometry over the course of the outburst, varying mass-transfer rates, or alternative outburst mechanisms \citep{patruno,goodwin2020b,gasealahwe}. These known dynamic changes within a single outburst obscure a conclusive interpretation of our inner disk radii obtained from the reflection component of the model. More observations are needed to further probe how the accretion disk geometry changes over the course of a single outburst. 

The fluctuation in central energy of the 1 keV feature motivates further investigation. The drastic shift in the 2019 observations may be explained by thermonuclear bursts, changes in reflection, or changes in ionization due to irradiation of the disk. We plan to extend our analysis to all available \textit{NICER} observations for the 2019 and 2022 outbursts, which could provide further insight into this elusive source, possibly supported by multi-wavelength analysis.

\section{Conclusion}\label{sec:conclusion}
In this paper, we analyzed the combined broad-band \textit{NICER} and \textit{NuSTAR} spectrum of SAX J1808 for the first time. In our detailed analysis, we identified several thermonuclear bursts occurring during the 2019 and 2022 observations performed by \textit{NICER}. We also performed time-resolved spectral analysis of two thermonuclear bursts occurring during the 2019 \textit{NuSTAR} observations. Our spectral modeling provides evidence of reflection during outbursts and constrains the inclination between 69-73\degree. We also detected the 1 keV feature in the spectra of both outbursts, at 0.90 keV in 2019 and 0.99 keV in 2022. This, along with the analyses of \cite{casten23}, \cite{bult}, and \cite{sharma23}, shows that this line varies greatly in central energy during the 2019 outburst. To the best of our knowledge, we are the first to analyze spectra from the 2022 outburst and the first to identify the 1 keV feature in this outburst. In a subsequent paper, we intend to continue our investigation of the \textit{NICER} data, with particular focus on this 1 keV feature. The constrained parameters from our broad-band analysis will aid in this future modeling.

\section*{Acknowledgments}\label{sec:acknowledgment}
We thank Dr. A. Fabian, Dr. A. Lohfink, Dr. T. Enoto, and the anonymous referee for their useful advice and comments. We also thank Mr. E. Bruce for his continuing technical support. Lastly, we would like to thank the Montana Space Grant Consortium for funding this research in part. This research has made use of data and/or software provided by the High Energy Astrophysics Science Archive Research Center (HEASARC), which is a service of the Astrophysics Science Division at NASA/GSFC. Computational efforts were performed on the Tempest High Performance Computing System, operated and supported by University Information Technology Research Cyberinfrastructure (RRID:SCR\textunderscore026229) at Montana State University.

\bibliographystyle{aasjournal}
\bibliography{bruceetal25}

@ARTICLE{bult,
       author = {{Bult}, Peter and {Jaisawal}, Gaurava K. and {G{\"u}ver}, Tolga and {Strohmayer}, Tod E. and {Altamirano}, Diego and {Arzoumanian}, Zaven and {Ballantyne}, David R. and {Chakrabarty}, Deepto and {Chenevez}, J{\'e}r{\^o}me and {Gendreau}, Keith C. and {Guillot}, Sebastien and {Ludlam}, Renee M.},
        title = "{A NICER Thermonuclear Burst from the Millisecond X-Ray Pulsar SAX J1808.4-3658}",
      journal = {\apjl},
         year = 2019,
        month = nov,
       volume = {885},
       number = {1},
          eid = {L1},
        pages = {L1},
          doi = {10.3847/2041-8213/ab4ae1},
archivePrefix = {arXiv},
       eprint = {1909.03595},
 primaryClass = {astro-ph.HE},
       adsurl = {https://ui.adsabs.harvard.edu/abs/2019ApJ...885L...1B}
}

@ARTICLE{disalvo,
       author = {{Di Salvo}, T. and {Sanna}, A. and {Burderi}, L. and {Papitto}, A. and {Iaria}, R. and {Gambino}, A.~F. and {Riggio}, A.},
        title = "{NuSTAR and XMM-Newton broad-band spectrum of SAX J1808.4-3658 during its latest outburst in 2015}",
      journal = {\mnras},
         year = 2019,
        month = feb,
       volume = {483},
       number = {1},
        pages = {767-779},
          doi = {10.1093/mnras/sty2974},
archivePrefix = {arXiv},
       eprint = {1811.00940},
 primaryClass = {astro-ph.HE},
       adsurl = {https://ui.adsabs.harvard.edu/abs/2019MNRAS.483..767D}
}

@ARTICLE{galloway,
       author = {{Galloway}, Duncan K. and {Cumming}, Andrew},
        title = "{Helium-rich Thermonuclear Bursts and the Distance to the Accretion-powered Millisecond Pulsar SAX J1808.4-3658}",
      journal = {\apj},
         year = 2006,
        month = nov,
       volume = {652},
       number = {1},
        pages = {559-568},
          doi = {10.1086/507598},
archivePrefix = {arXiv},
       eprint = {astro-ph/0607213},
 primaryClass = {astro-ph},
       adsurl = {https://ui.adsabs.harvard.edu/abs/2006ApJ...652..559G}
}

@ARTICLE{garcia,
       author = {{Garc{\'\i}a}, J. and {Dauser}, T. and {Lohfink}, A. and {Kallman}, T.~R. and {Steiner}, J.~F. and {McClintock}, J.~E. and {Brenneman}, L. and {Wilms}, J. and {Eikmann}, W. and {Reynolds}, C.~S. and {Tombesi}, F.},
        title = "{Improved Reflection Models of Black Hole Accretion Disks: Treating the Angular Distribution of X-Rays}",
      journal = {\apj},
         year = 2014,
        month = feb,
       volume = {782},
       number = {2},
          eid = {76},
        pages = {76},
          doi = {10.1088/0004-637X/782/2/76},
archivePrefix = {arXiv},
       eprint = {1312.3231},
 primaryClass = {astro-ph.HE},
       adsurl = {https://ui.adsabs.harvard.edu/abs/2014ApJ...782...76G}
}

@ARTICLE{fabian2000,
       author = {{Fabian}, A.~C. and {Iwasawa}, K. and {Reynolds}, C.~S. and {Young}, A.~J.},
        title = "{Broad Iron Lines in Active Galactic Nuclei}",
      journal = {\pasp},
         year = 2000,
        month = sep,
       volume = {112},
       number = {775},
        pages = {1145-1161},
          doi = {10.1086/316610},
archivePrefix = {arXiv},
       eprint = {astro-ph/0004366},
 primaryClass = {astro-ph},
       adsurl = {https://ui.adsabs.harvard.edu/abs/2000PASP..112.1145F}
}

@ARTICLE{wijnands,
       author = {{Wijnands}, Rudy and {van der Klis}, Michiel},
        title = "{The Broadband Power Spectrum of SAX J1808.4-3658}",
      journal = {\apjl},
         year = 1998,
        month = nov,
       volume = {507},
       number = {1},
        pages = {L63-L66},
          doi = {10.1086/311676},
archivePrefix = {arXiv},
       eprint = {astro-ph/9808303},
 primaryClass = {astro-ph},
       adsurl = {https://ui.adsabs.harvard.edu/abs/1998ApJ...507L..63W}
}

@ARTICLE{papitto09,
       author = {{Papitto}, A. and {Di Salvo}, T. and {D'A{\`\i}}, A. and {Iaria}, R. and {Burderi}, L. and {Riggio}, A. and {Menna}, M.~T. and {Robba}, N.~R.},
        title = "{XMM-Newton detects a relativistically broadened iron line in the spectrum of the ms X-ray pulsar SAX J1808.4-3658}",
      journal = {\aap},
         year = 2009,
        month = jan,
       volume = {493},
       number = {3},
        pages = {L39-L43},
          doi = {10.1051/0004-6361:200811401},
archivePrefix = {arXiv},
       eprint = {0812.1149},
 primaryClass = {astro-ph},
       adsurl = {https://ui.adsabs.harvard.edu/abs/2009A&A...493L..39P}
}

@ARTICLE{bildsten,
       author = {{Bildsten}, Lars and {Chakrabarty}, Deepto},
        title = "{A Brown Dwarf Companion for the Accreting Millisecond Pulsar SAX J1808.4-3658}",
      journal = {\apj},
         year = 2001,
        month = aug,
       volume = {557},
       number = {1},
        pages = {292-296},
          doi = {10.1086/321633},
archivePrefix = {arXiv},
       eprint = {astro-ph/0104153},
 primaryClass = {astro-ph},
       adsurl = {https://ui.adsabs.harvard.edu/abs/2001ApJ...557..292B}
}

@ARTICLE{goodwin2019,
       author = {{Goodwin}, A.~J. and {Galloway}, D.~K. and {Heger}, A. and {Cumming}, A. and {Johnston}, Z.},
        title = "{A Bayesian approach to matching thermonuclear X-ray burst observations with models}",
      journal = {\mnras},
         year = 2019,
        month = dec,
       volume = {490},
       number = {2},
        pages = {2228-2240},
          doi = {10.1093/mnras/stz2638},
archivePrefix = {arXiv},
       eprint = {1907.00996},
 primaryClass = {astro-ph.HE},
       adsurl = {https://ui.adsabs.harvard.edu/abs/2019MNRAS.490.2228G}
}

@ARTICLE{galloway08,
       author = {{Galloway}, Duncan K. and {Muno}, Michael P. and {Hartman}, Jacob M. and {Psaltis}, Dimitrios and {Chakrabarty}, Deepto},
        title = "{Thermonuclear (Type I) X-Ray Bursts Observed by the Rossi X-Ray Timing Explorer}",
      journal = {\apjs},
         year = 2008,
        month = dec,
       volume = {179},
       number = {2},
        pages = {360-422},
          doi = {10.1086/592044},
archivePrefix = {arXiv},
       eprint = {astro-ph/0608259},
 primaryClass = {astro-ph},
       adsurl = {https://ui.adsabs.harvard.edu/abs/2008ApJS..179..360G}
}

@ARTICLE{straaten,
       author = {{van Straaten}, Steve and {van der Klis}, Michiel and {Wijnands}, Rudy},
        title = "{The aperiodic timing behaviour of the accretion-driven millisecond pulsar SAX J1808.4-3658}",
      journal = {Nuclear Physics B Proceedings Supplements},
         year = 2004,
        month = jun,
       volume = {132},
        pages = {664-667},
          doi = {10.1016/j.nuclphysbps.2004.04.118},
archivePrefix = {arXiv},
       eprint = {astro-ph/0309345},
 primaryClass = {astro-ph},
       adsurl = {https://ui.adsabs.harvard.edu/abs/2004NuPhS.132..664V}
}

@ARTICLE{rupen,
       author = {{Rupen}, M.~P. and {Dhawan}, V. and {Mioduszewski}, A.~J.},
        title = "{Radio detections of SAX J1808.4-3658}",
      journal = {The Astronomer's Telegram},
         year = 2005,
        month = jun,
       volume = {524},
        pages = {1},
       adsurl = {https://ui.adsabs.harvard.edu/abs/2005ATel..524....1R}
}

@ARTICLE{markwardt08,
       author = {{Markwardt}, C.~B. and {Swank}, J.~H.},
        title = "{RXTE PCA Detects the Millisecond Pulsar SAX J1808.4-3658 in Outburst}",
      journal = {The Astronomer's Telegram},
         year = 2008,
        month = sep,
       volume = {1728},
        pages = {1},
       adsurl = {https://ui.adsabs.harvard.edu/abs/2008ATel.1728....1M}
}

@ARTICLE{altamirano,
       author = {{Altamirano}, D. and {Kaur}, R. and {Degenaar}, N. and {Wijnands}, R. and {Yang}, Y. and {Armas-Padilla}, M. and {Strohmayer},, T. and {Markwardt}, C.},
        title = "{Possible outburst in neutron star LMXB SAX J1806.5-2215}",
      journal = {The Astronomer's Telegram},
         year = 2011,
        month = feb,
       volume = {3193},
        pages = {1},
       adsurl = {https://ui.adsabs.harvard.edu/abs/2011ATel.3193....1A}
}

@ARTICLE{sanna,
       author = {{Sanna}, A. and {Pintore}, F. and {Riggio}, A. and {D'Ai}, A. and {Di Salvo}, T. and {Burderi}, L. and {Iaria}, R. and {Scarano}, F. and {Segreto}, A.},
        title = "{SWIFT/XRT confirmed the new outburst of SAX J1808.4-3658}",
      journal = {The Astronomer's Telegram},
         year = 2015,
        month = apr,
       volume = {7371},
        pages = {1},
       adsurl = {https://ui.adsabs.harvard.edu/abs/2015ATel.7371....1S}
}

@ARTICLE{zand,
       author = {{in 't Zand}, J.~J.~M. and {Heise}, J. and {Muller}, J.~M. and {Bazzano}, A. and {Cocchi}, M. and {Natalucci}, L. and {Ubertini}, P.},
        title = "{Discovery of the X-ray transient SAX J1808.4-3658, a likely low-mass X-ray binary}",
      journal = {\aap},
         year = 1998,
        month = mar,
       volume = {331},
        pages = {L25-L28},
          doi = {10.48550/arXiv.astro-ph/9802098},
archivePrefix = {arXiv},
       eprint = {astro-ph/9802098},
 primaryClass = {astro-ph},
       adsurl = {https://ui.adsabs.harvard.edu/abs/1998A&A...331L..25I}
}

@ARTICLE{markwardt02,
       author = {{Markwardt}, C.~B. and {Miller}, J.~M. and {Wijnands}, R.},
        title = "{New outburst of SAX J1808.4-3658}",
      journal = {The Astronomer's Telegram},
         year = 2002,
        month = oct,
       volume = {110},
        pages = {1},
       adsurl = {https://ui.adsabs.harvard.edu/abs/2002ATel..110....1M}
}

@ARTICLE{chakrabarty,
       author = {{Chakrabarty}, D. and {Morgan}, E.~H.},
        title = "{SAX J1808.4-3658 = XTE J1808-369}",
      journal = {\iaucirc},
         year = 1998,
        month = apr,
       volume = {6877},
        pages = {2},
       adsurl = {https://ui.adsabs.harvard.edu/abs/1998IAUC.6877....2C}
}

@ARTICLE{speicher2022,
       author = {{Speicher}, J. and {Ballantyne}, D.~R. and {Fragile}, P.~C.},
        title = "{Evolution of accretion disc reflection spectra due to a Type I X-ray burst}",
      journal = {\mnras},
         year = 2022,
        month = jan,
       volume = {509},
       number = {2},
        pages = {1736-1744},
          doi = {10.1093/mnras/stab3087},
archivePrefix = {arXiv},
       eprint = {2110.11931},
 primaryClass = {astro-ph.HE},
       adsurl = {https://ui.adsabs.harvard.edu/abs/2022MNRAS.509.1736S}
}

@ARTICLE{kuul,
       author = {{Kuulkers}, E. and {den Hartog}, P.~R. and {in't Zand}, J.~J.~M. and {Verbunt}, F.~W.~M. and {Harris}, W.~E. and {Cocchi}, M.},
        title = "{Photospheric radius expansion X-ray bursts as standard candles}",
      journal = {\aap},
         year = 2003,
        month = feb,
       volume = {399},
        pages = {663-680},
          doi = {10.1051/0004-6361:20021781},
archivePrefix = {arXiv},
       eprint = {astro-ph/0212028},
 primaryClass = {astro-ph},
       adsurl = {https://ui.adsabs.harvard.edu/abs/2003A&A...399..663K}
}

@ARTICLE{chakraborty,
       author = {{Chakraborty}, Priyanka and {Ferland}, Gary and {Fabian}, Andrew and {Sarkar}, Arnab and {Ludlam}, Renee and {Bianchi}, Stefano and {Hall}, Hayden and {Kosec}, Peter},
        title = "{Physics of 1 keV line in X-ray binaries}",
      journal = {arXiv e-prints},
         year = 2024,
        month = jul,
          eid = {arXiv:2407.02360},
        pages = {arXiv:2407.02360},
          doi = {10.48550/arXiv.2407.02360},
archivePrefix = {arXiv},
       eprint = {2407.02360},
 primaryClass = {astro-ph.HE},
       adsurl = {https://ui.adsabs.harvard.edu/abs/2024arXiv240702360C}
}

@ARTICLE{ludlam2018,
       author = {{Ludlam}, R.~M. and {Miller}, J.~M. and {Arzoumanian}, Z. and {Bult}, P.~M. and {Cackett}, E.~M. and {Chakrabarty}, D. and {Dauser}, T. and {Enoto}, T. and {Fabian}, A.~C. and {Garc{\'\i}a}, J.~A. and et al.},
        title = "{Detection of Reflection Features in the Neutron Star Low-mass X-Ray Binary Serpens X-1 with NICER}",
      journal = {\apjl},
         year = 2018,
        month = may,
       volume = {858},
       number = {1},
          eid = {L5},
        pages = {L5},
          doi = {10.3847/2041-8213/aabee6},
archivePrefix = {arXiv},
       eprint = {1804.10214},
 primaryClass = {astro-ph.HE},
       adsurl = {https://ui.adsabs.harvard.edu/abs/2018ApJ...858L...5L}
}

@ARTICLE{romanova,
       author = {{Romanova}, Marina M. and {Owocki}, Stanley P.},
        title = "{Accretion, Outflows, and Winds of Magnetized Stars}",
      journal = {\ssr},
         year = 2015,
        month = oct,
       volume = {191},
       number = {1-4},
        pages = {339-389},
          doi = {10.1007/s11214-015-0200-9},
archivePrefix = {arXiv},
       eprint = {1605.04979},
 primaryClass = {astro-ph.SR},
       adsurl = {https://ui.adsabs.harvard.edu/abs/2015SSRv..191..339R}
}

@ARTICLE{chen,
       author = {{Chen}, Wan and {Shrader}, C.~R. and {Livio}, Mario},
        title = "{The Properties of X-Ray and Optical Light Curves of X-Ray Novae}",
      journal = {\apj},
         year = 1997,
        month = dec,
       volume = {491},
       number = {1},
        pages = {312-338},
          doi = {10.1086/304921},
archivePrefix = {arXiv},
       eprint = {astro-ph/9707138},
 primaryClass = {astro-ph},
       adsurl = {https://ui.adsabs.harvard.edu/abs/1997ApJ...491..312C}
}

@ARTICLE{hall25,
       author = {{Hall}, H. and {Ludlam}, R.~M. and {Miller}, J.~M. and {Fabian}, A.~C. and {Tomsick}, J.~A. and {Coley}, J. and {Garc{\'\i}a}, J.~A. and {Coughenour}, B.~M.},
        title = "{Simultaneous NICER and NuSTAR Observations of the Neutron Star Low-mass X-Ray Binary Serpens X-1}",
      journal = {\apj},
         year = 2025,
        month = feb,
       volume = {980},
       number = {2},
          eid = {234},
        pages = {234},
          doi = {10.3847/1538-4357/adaeaa},
archivePrefix = {arXiv},
       eprint = {2501.17137},
 primaryClass = {astro-ph.HE},
       adsurl = {https://ui.adsabs.harvard.edu/abs/2025ApJ...980..234H}
}

@ARTICLE{sharma23,
       author = {{Sharma}, Rahul and {Sanna}, Andrea and {Beri}, Aru},
        title = "{AstroSat observation of the accreting millisecond X-ray pulsar SAX J1808.4-3658 during its 2019 outburst}",
      journal = {\mnras},
         year = 2023,
        month = mar,
       volume = {519},
       number = {3},
        pages = {3811-3818},
          doi = {10.1093/mnras/stac3779},
archivePrefix = {arXiv},
       eprint = {2212.10088},
 primaryClass = {astro-ph.HE},
       adsurl = {https://ui.adsabs.harvard.edu/abs/2023MNRAS.519.3811S}
}

@ARTICLE{casten23,
       author = {{Casten}, Sierra and {Strohmayer}, Tod E. and {Bult}, Peter},
        title = "{Hydrogen-triggered X-Ray Bursts from SAX J1808.4-3658? The Onset of Nuclear Burning}",
      journal = {\apj},
         year = 2023,
        month = may,
       volume = {948},
       number = {2},
          eid = {117},
        pages = {117},
          doi = {10.3847/1538-4357/acc24f},
archivePrefix = {arXiv},
       eprint = {2304.00104},
 primaryClass = {astro-ph.HE},
       adsurl = {https://ui.adsabs.harvard.edu/abs/2023ApJ...948..117C}
}

@ARTICLE{wilms,
       author = {{Wilms}, J. and {Allen}, A. and {McCray}, R.},
        title = "{On the Absorption of X-Rays in the Interstellar Medium}",
      journal = {\apj},
         year = 2000,
        month = oct,
       volume = {542},
       number = {2},
        pages = {914-924},
          doi = {10.1086/317016},
archivePrefix = {arXiv},
       eprint = {astro-ph/0008425},
 primaryClass = {astro-ph},
       adsurl = {https://ui.adsabs.harvard.edu/abs/2000ApJ...542..914W}
}

@ARTICLE{zycki,
       author = {{{\.Z}ycki}, Piotr T. and {Done}, Chris and {Smith}, David A.},
        title = "{The 1989 May outburst of the soft X-ray transient GS 2023+338 (V404 Cyg)}",
      journal = {\mnras},
         year = 1999,
        month = nov,
       volume = {309},
       number = {3},
        pages = {561-575},
          doi = {10.1046/j.1365-8711.1999.02885.x},
archivePrefix = {arXiv},
       eprint = {astro-ph/9904304},
 primaryClass = {astro-ph},
       adsurl = {https://ui.adsabs.harvard.edu/abs/1999MNRAS.309..561Z}
}

@ARTICLE{zdziarski,
       author = {{Zdziarski}, A.~A. and {Johnson}, W.~N. and {Magdziarz}, P.},
        title = "{Broad-band {\ensuremath{\gamma}}-ray and X-ray spectra of NGC 4151 and their implications for physical processes and geometry.}",
      journal = {\mnras},
         year = 1996,
        month = nov,
       volume = {283},
       number = {1},
        pages = {193-206},
          doi = {10.1093/mnras/283.1.193},
archivePrefix = {arXiv},
       eprint = {astro-ph/9607015},
 primaryClass = {astro-ph},
       adsurl = {https://ui.adsabs.harvard.edu/abs/1996MNRAS.283..193Z}
}

@ARTICLE{harrison,
       author = {{Harrison}, Fiona A. and {Craig}, William W. and {Christensen}, Finn E. and {Hailey}, Charles J. and {Zhang}, William W. and {Boggs}, Steven E. and {Stern}, Daniel and {Cook}, W. Rick and {Forster}, Karl and {Giommi}, Paolo and et al.},
        title = "{The Nuclear Spectroscopic Telescope Array (NuSTAR) High-energy X-Ray Mission}",
      journal = {\apj},
         year = 2013,
        month = jun,
       volume = {770},
       number = {2},
          eid = {103},
        pages = {103},
          doi = {10.1088/0004-637X/770/2/103},
archivePrefix = {arXiv},
       eprint = {1301.7307},
 primaryClass = {astro-ph.IM},
       adsurl = {https://ui.adsabs.harvard.edu/abs/2013ApJ...770..103H}
}

@INPROCEEDINGS{blackburn,
       author = {{Blackburn}, J.~K.},
        title = "{FTOOLS: A FITS Data Processing and Analysis Software Package}",
    booktitle = {Astronomical Data Analysis Software and Systems IV},
         year = 1995,
       editor = {{Shaw}, R.~A. and {Payne}, H.~E. and {Hayes}, J.~J.~E.},
       series = {Astronomical Society of the Pacific Conference Series},
       volume = {77},
        month = jan,
        pages = {367},
       adsurl = {https://ui.adsabs.harvard.edu/abs/1995ASPC...77..367B}
}

@INPROCEEDINGS{arnaud,
       author = {{Arnaud}, K.~A.},
        title = "{XSPEC: The First Ten Years}",
    booktitle = {Astronomical Data Analysis Software and Systems V},
         year = 1996,
       editor = {{Jacoby}, George H. and {Barnes}, Jeannette},
       series = {Astronomical Society of the Pacific Conference Series},
       volume = {101},
        month = jan,
        pages = {17},
       adsurl = {https://ui.adsabs.harvard.edu/abs/1996ASPC..101...17A}
}

@ARTICLE{white,
       author = {{White}, N.~E. and {Peacock}, A. and {Hasinger}, G. and {Mason}, K.~O. and {Manzo}, G. and {Taylor}, B.~G. and {Branduardi-Raymont}, G.},
        title = "{A study of the continuum and iron K line emission from low-mass X-raybinaries.}",
      journal = {\mnras},
         year = 1986,
        month = jan,
       volume = {218},
        pages = {129-138},
          doi = {10.1093/mnras/218.1.129},
       adsurl = {https://ui.adsabs.harvard.edu/abs/1986MNRAS.218..129W}
}

@BOOK{osterbrock,
       author = {{Osterbrock}, Donald E. and {Ferland}, Gary J.},
        title = "{Astrophysics of gaseous nebulae and active galactic nuclei}",
         year = 2006,
        publisher = "{University Science Books}", 
       adsurl = {https://ui.adsabs.harvard.edu/abs/2006agna.book.....O}
}

@ARTICLE{goodwin2020b,
       author = {{Goodwin}, A.~J. and {Russell}, D.~M. and {Galloway}, D.~K. and {Baglio}, M.~C. and {Parikh}, A.~S. and {Buckley}, D.~A.~H. and {Homan}, J. and {Bramich}, D.~M. and {in't Zand}, J.~J.~M. and {Heinke}, C.~O. and et al.},
        title = "{Enhanced optical activity 12 d before X-ray activity, and a 4 d X-ray delay during outburst rise, in a low-mass X-ray binary}",
      journal = {\mnras},
         year = 2020,
        month = nov,
       volume = {498},
       number = {3},
        pages = {3429-3439},
          doi = {10.1093/mnras/staa2588},
archivePrefix = {arXiv},
       eprint = {2006.02872},
 primaryClass = {astro-ph.HE},
       adsurl = {https://ui.adsabs.harvard.edu/abs/2020MNRAS.498.3429G}
}

@ARTICLE{bult2015,
       author = {{Bult}, Peter and {van der Klis}, Michiel},
        title = "{The Aperiodic X-Ray Variability of the Accreting Millisecond Pulsar SAX J1808.4-3658}",
      journal = {\apj},
         year = 2015,
        month = jun,
       volume = {806},
       number = {1},
          eid = {90},
        pages = {90},
          doi = {10.1088/0004-637X/806/1/90},
archivePrefix = {arXiv},
       eprint = {1505.00596},
 primaryClass = {astro-ph.HE},
       adsurl = {https://ui.adsabs.harvard.edu/abs/2015ApJ...806...90B}
}

@ARTICLE{baglio2020,
       author = {{Baglio}, M.~C. and {Russell}, D.~M. and {Crespi}, S. and {Covino}, S. and {Johar}, A. and {Homan}, J. and {Bramich}, D.~M. and {Saikia}, P. and {Campana}, S. and {D'Avanzo}, P. and et al.},
        title = "{Probing Jet Launching in Neutron Star X-Ray Binaries: The Variable and Polarized Jet of SAX J1808.4-3658}",
      journal = {\apj},
         year = 2020,
        month = dec,
       volume = {905},
       number = {2},
          eid = {87},
        pages = {87},
          doi = {10.3847/1538-4357/abc685},
archivePrefix = {arXiv},
       eprint = {2010.15176},
 primaryClass = {astro-ph.HE},
       adsurl = {https://ui.adsabs.harvard.edu/abs/2020ApJ...905...87B}
}

@ARTICLE{patruno,
       author = {{Patruno}, A. and {Maitra}, D. and {Curran}, P.~A. and {D'Angelo}, C. and {Fridriksson}, J.~K. and {Russell}, D.~M. and {Middleton}, M. and {Wijnands}, R.},
        title = "{The Reflares and Outburst Evolution in the Accreting Millisecond Pulsar SAX J1808.4-3658: A Disk Truncated Near Co-Rotation?}",
      journal = {\apj},
         year = 2016,
        month = feb,
       volume = {817},
       number = {2},
          eid = {100},
        pages = {100},
          doi = {10.3847/0004-637X/817/2/100},
archivePrefix = {arXiv},
       eprint = {1504.05048},
 primaryClass = {astro-ph.HE},
       adsurl = {https://ui.adsabs.harvard.edu/abs/2016ApJ...817..100P}
}

@ARTICLE{gasealahwe,
       author = {{Gasealahwe}, K.~V.~S. and {Monageng}, I.~M. and {Fender}, R.~P. and {Woudt}, P.~A. and {Motta}, S.~E. and {van den Eijnden}, J. and {Williams}, D.~R.~A. and {Heywood}, I. and {Bloemen}, S. and {Groot}, P.~J. and et al.},
        title = "{The 2019 outburst of AMXP SAX J1808.4-3658 and radio follow up of MAXI J0911-655 and XTE J1701-462}",
      journal = {\mnras},
         year = 2023,
        month = may,
       volume = {521},
       number = {2},
        pages = {2806-2813},
          doi = {10.1093/mnras/stad649},
archivePrefix = {arXiv},
       eprint = {2302.13899},
 primaryClass = {astro-ph.HE},
       adsurl = {https://ui.adsabs.harvard.edu/abs/2023MNRAS.521.2806G}
}

@ARTICLE{asquini,
       author = {{Asquini}, L. and {Baglio}, M.~C. and {Campana}, S. and {D'Avanzo}, P. and {Miraval Zanon}, A. and {Alabarta}, K. and {Russell}, D.~M. and {Bramich}, D.~M.},
        title = "{Lack of emission lines in the optical spectra of SAX J1808.4{\textendash}3658 during reflaring of the 2019 outburst}",
      journal = {\aap},
         year = 2024,
        month = dec,
       volume = {692},
          eid = {A16},
        pages = {A16},
          doi = {10.1051/0004-6361/202450816},
archivePrefix = {arXiv},
       eprint = {2411.04828},
 primaryClass = {astro-ph.HE},
       adsurl = {https://ui.adsabs.harvard.edu/abs/2024A&A...692A..16A}
}

@ARTICLE{wijnands2001,
       author = {{Wijnands}, Rudy and {M{\'e}ndez}, Mariano and {Markwardt}, Craig and {van der Klis}, Michiel and {Chakrabarty}, Deepto and {Morgan}, Ed},
        title = "{The Erratic Luminosity Behavior of SAX J1808.4-3658 during Its 2000 Outburst}",
      journal = {The Astrophysical Journal},
         year = 2001,
        month = oct,
       volume = {560},
       number = {2},
        pages = {892-896},
          doi = {10.1086/323073},
archivePrefix = {arXiv},
       eprint = {astro-ph/0105446},
 primaryClass = {astro-ph},
       adsurl = {https://ui.adsabs.harvard.edu/abs/2001ApJ...560..892W}
}

@ARTICLE{patruno2009,
       author = {{Patruno}, Alessandro and {Watts}, Anna and {Klein Wolt}, Marc and {Wijnands}, Rudy and {van der Klis}, Michiel},
        title = "{1 Hz Flaring in SAX J1808.4-3658: Flow Instabilities near the Propeller Stage}",
      journal = {The Astrophysical Journal},
         year = 2009,
        month = dec,
       volume = {707},
       number = {2},
        pages = {1296-1309},
          doi = {10.1088/0004-637X/707/2/1296},
archivePrefix = {arXiv},
       eprint = {0904.0560},
 primaryClass = {astro-ph.HE},
       adsurl = {https://ui.adsabs.harvard.edu/abs/2009ApJ...707.1296P}
}

@ARTICLE{pappitto2013,
       author = {{Papitto}, A. and {D'A{\`\i}}, A. and {Di Salvo}, T. and {Egron}, E. and {Bozzo}, E. and {Burderi}, L. and {Iaria}, R. and {Riggio}, A. and {Menna}, M.~T.},
        title = "{The accretion flow to the intermittent accreting millisecond pulsar, HETE J1900.1-2455, as observed by XMM-Newton and RXTE}",
      journal = {Monthly Notices of the Royal Astronomical Society},
         year = 2013,
        month = mar,
       volume = {429},
       number = {4},
        pages = {3411-3422},
          doi = {10.1093/mnras/sts605},
archivePrefix = {arXiv},
       eprint = {1212.2532},
 primaryClass = {astro-ph.HE},
       adsurl = {https://ui.adsabs.harvard.edu/abs/2013MNRAS.429.3411P}
}

\begin{figure*}[h]
  \centering
  \begin{subfigure}{0.45\textwidth}
    \centering
    \includegraphics[width=\linewidth]{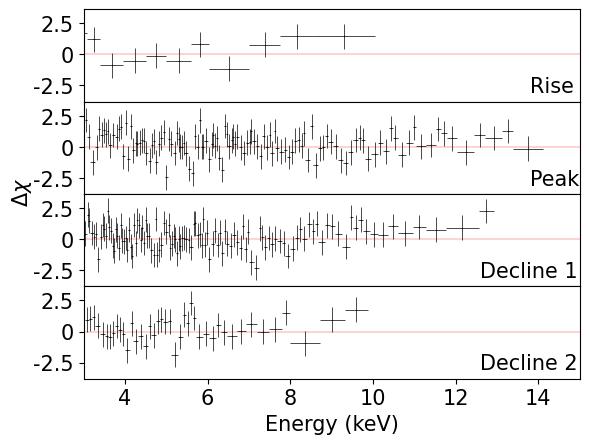}
    \caption{Burst 1, starting at t$_{0}$=32280 s}
    \label{appendix:burst1resid}
  \end{subfigure}
  \quad
  \begin{subfigure}{0.45\textwidth}
    \centering
    \includegraphics[width=\linewidth]{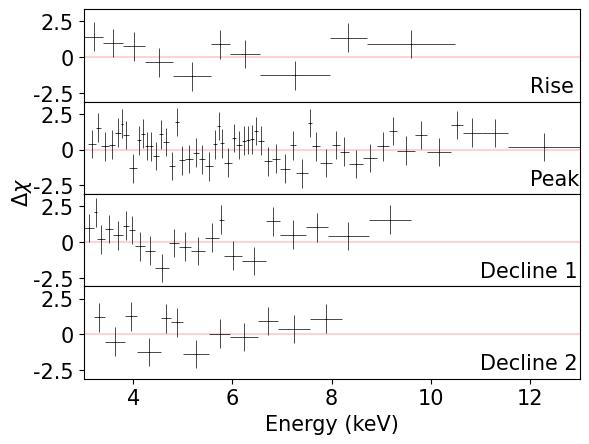}
    \caption{Burst 2, starting at t$_{0}$=72010 s}
    \label{appendix:burst2resid}
  \end{subfigure}
  \caption{ Residuals in units of sigma for each section of the modeled spectra during the two thermonuclear bursts. Top: rise. Top middle: peak. Bottom middle: decline 1. Bottom: decline 2.}
  \label{fig:burstresid}
\end{figure*}

\appendix\label{sec:appendix}
The resuiduals of the modeled spectra during the two thermonuclear bursts occurring during the 2019 NuSTAR observation are show in Figure \ref{fig:burstresid}. Each section of the burst was modeled with photoelectric absorption (including interstellar oxygen and iron abundance) and blackbody components.

\end{document}